**Gamma Group**

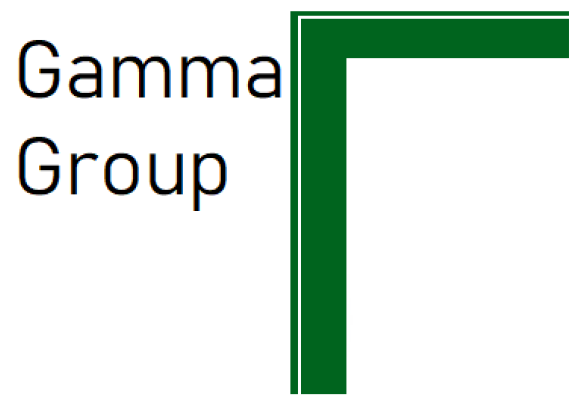

# Decision Forest Based EMG Signal Classification with Low Volume Dataset Augmented with Random Variance Gaussian Noise

Tekin Gunasar, Alexandra Rekesh, Atul Nair, Penelope King,
Anastasiya Markova, Jiaqi Zhang, and Isabel Tate

2022

# Decision Forest Based EMG Signal Classification with Low Volume Dataset Augmented with Random Variance Gaussian Noise

*Abstract*—**Electromyography signals can be used as training data by machine learning models to classify various gestures. We seek to produce a model that can classify six different hand gestures with a limited number of samples that generalizes well to a wider audience while comparing the effect of our feature extraction results on model accuracy to other more conventional methods such as the use of AR parameters on a sliding window across the channels of a signal. We appeal to a set of more elementary methods such as the use of random bounds on a signal, but desire to show the power these methods can carry in an online setting where EMG classification is being conducted, as opposed to more complicated methods such as the use of the Fourier Transform. To augment our limited training data, we used a standard technique, known as jitter, where random noise is added to each observation in a channel wise manner. Once all datasets were produced using the above methods, we performed a grid search with Random Forest and XGBoost to ultimately create a high accuracy model. For human computer interface purposes, high accuracy classification of EMG signals is of particular importance to their functioning and given the difficulty and cost of amassing any sort of biomedical data in a high volume, it is valuable to have techniques that can work with a low amount of high-quality samples with less expensive feature extraction methods that can reliably be carried out in an online application.**

## I.    Introduction

Human computer interfaces are becoming increasingly relevant in our modern era. On one side, there is a possibility of helping disabled persons conducting their day to day lives through technologies including word processing programs and functional prosthetic limbs for amputees. While on the other hand, the idea of human computer interfaces arises naturally in many new technologies such as virtual reality, or really any other system that would require a continuous classification of bioelectrical signals [1]. Particularly, an EMG signal is an electrical current generated from a contracting muscle. To measure these signals, electrodes are typically placed onto the surface of the skin, thus this is dubbed as an sEMG(Surface Electromyography). While this signal is measured, it travels through other tissue as well, and noise from other tissue near the one being measured is added onto the target signal, making each signal measured, even if from the same tissue, inherently very variable[2]. One possible solution is to instead perform an intramuscular electromyography, where electrodes are directly inserted into muscle tissue which would cause the noise in the signal from our target muscle to be considerably less, but of course this is a rather uncomfortable procedure and is not a viable option for the creation of large data sets. Thus, EMG signals are naturally very noisy and complex, and depending on how many channels are being measured can be very high dimensional. Many related models that aid to classify these signals are neural network based [1]. While neural networks are known to generalize well for an appropriate amount of training data, they can be rather costly and difficult to train, and may not be the best option for scenarios with limited samples, even when considering proper data augmentation. To make these signals simpler to work with, a variety of denoising techniques can be applied, these include wavelet transform, and by extension the Fourier transform, based denoising[3], different filters such as Butterworth and Wiener filters[4]; and even neural network-based approaches[5]. While neural network based methods are popular in real time classification of EMG signals, tree-based models also yield promising results, as was demonstrated in [6] where decision tree algorithms combined with multiscale principal component analysis (MSPCA) for noise reduction yielded a promising 96.67% classification accuracy.

However, we wish to stray away from evaluating the quality of a trained model by simply attempting to maximize a metric like accuracy, and rather take on a more general view of creating a model that can be deployed in practical applications. This is especially relevant for our purposes with a small dataset[16] in preventing overfitting on our training set and preserving the use of our model for any BCI where EMG classification is taking place in a more robust setting. Mukhopadhyay et al. [13] classify EMG signals corresponding to different upper limb movements with a DNN and achieve an impressive 98.88% accuracy, while their usage of Random Forest, with different features from our model, had an average accuracy of 91.78%. This is inline with our claims that the usage of more computationally expensive deep learning models can be avoided, in favor of more classical models, such as the previously mentioned Random Forest. Deep learning models perform rather well when there is sufficient data to make sure that that they do not learn the unique features to their training sets, but EMG data is rather difficult, and may be expensive to acquire in a high enough volume to make the usage of deep neural networks ideal. This is explored by Rehman et al. [14] where a CNN and SSAE-f was used to classify raw EMG signals, which also achieved high accuracy results, but the importance of more training data was emphasized for deep learning models to perform better "in the wild".

The rest of  this paper will be divided as following: We first cover the  specific six gestures we seek to classify and how we collected data, an overview of our feature extraction and data augmentation techniques, experimental results using these methods on Random Forest and XGBoost models, and concluding with a discussion of the implications of being able to classify bioelectric signals with limited samples, as well as what it means to be moving away from deep learning neural network based approaches for human computer interfaces.

## II.    Methods

### II.1 Data Collection

Data for six different hand gestures was collected: open fist, clenched fist, upward wrist rotation, downward wrist rotation, left wrist rotation, and right wrist rotation. Below is a panel of images of each of these gestures.

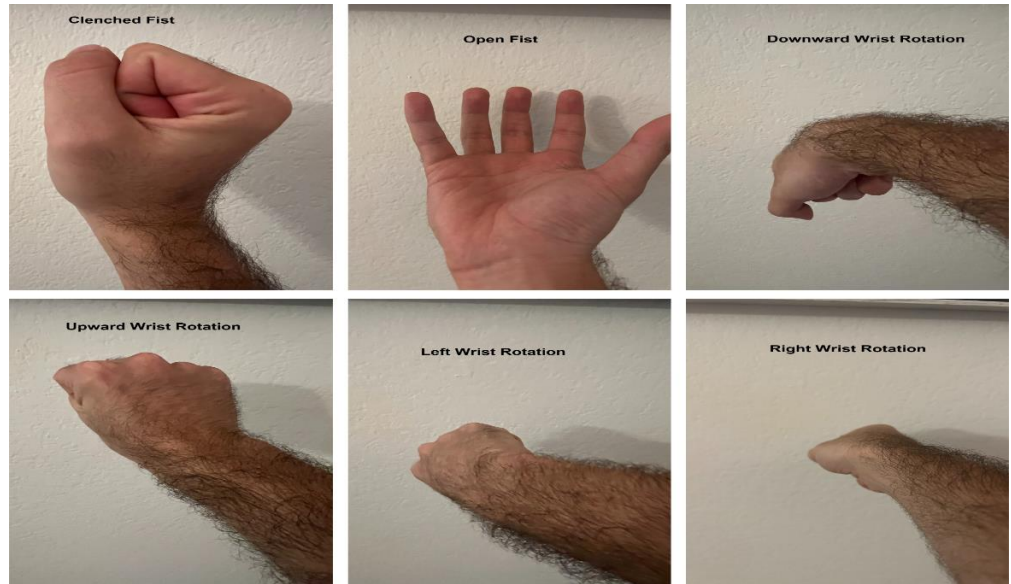

Fig 1. A diagram of the six different hand gestures recorded

Data was recorded on three different channels, and ten observations were recorded for each hand gesture, hence, a total of sixty observations. Each observation was recorded for 650ms, but the for the first 200ms the muscle was relaxed to prevent addition of noise from the transition of previous hand gestures into the next. Given our desire to continuously classify EMG signals, we will be applying an FIR filter. FIR filters have a linear phase shift which means we will have a constant delay that can be corrected for in a real time system. Additionally, a bandpass was taken from 0.1Hz to 125 Hz with a sampling rate of 125Hz, thus, by the Nyquist theorem the maximum of a processed signal is one half of the sampling rate. This means that all signals faster than 125Hz are likely aliased and can be discarded. Finally, for smoothing of data, a moving average over 50ms intervals was applied to each signal in a channel wise fashion. Below is an example of one of these observations after pre-processing.

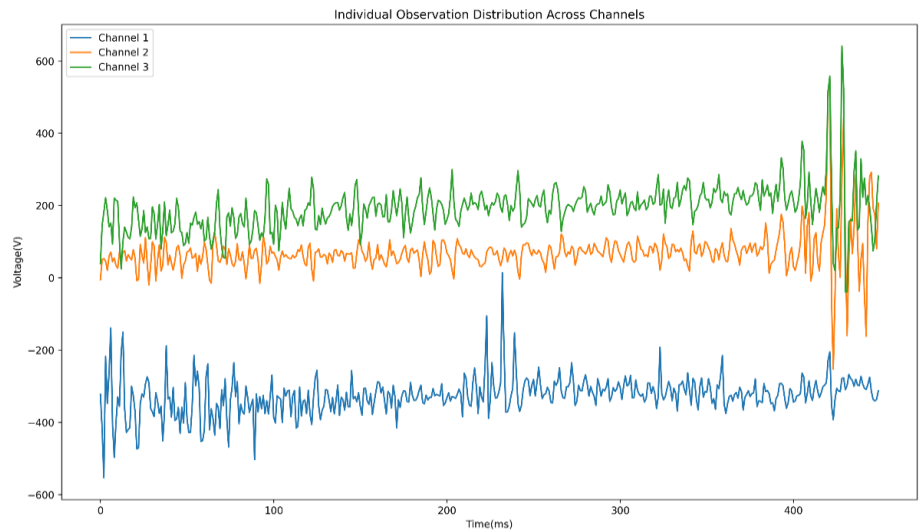

Fig 2. The third observation in the group of downward wrist rotations, where the first 200ms of no muscle activity cut off.

*II.2 Feature Extraction, Data Augmentation, and Model Choice*

We used two types of feature extraction techniques: using strictly the raw values of the signals, and modeling signals as an AR model and using the associated AR coefficients as features. Many more complex methods of feature extraction for EMG signals exist, such as integrated absolute value of third and second derivatives, integrated absolute log values, the more novel L-index, etc.;[9], however, these were not pursued, as this paper is not a survey of all feature extraction methods. EMG signals were first modeled as autoregressive moving average models by Graupe and Cline in 1975 and were demonstrated to be stationary within a short enough time interval[7]. However, given the overall non-stationarity of EMG signals, these methods were later improved upon by Zhou et al.[8] in 1986 and brought about the idea of using AR models to represent EMG signals.

To augment our data, we added Gaussian noise on each observation with mean zero, and the variance was randomly chosen from a Gamma distribution with shape parameter three, and scaling parameter two, which is a variation of the standard jitter technique [15]. Noise was added to each of the 60 observations 25 times, leading to 1500 total observations after augmentation. Each observation had noise added to it by a different Gaussian distribution to account for the interpersonal variation of recorded EMG signals. If noise was only added from one Gaussian distribution, the augmented signals would still be too similar and might not prove to be very effective in creating an applicable model with a small dataset.

We created one type of dataset associated with AR parameters where we first performed the previously mentioned augmentation technique. Each observation then was cropped in a channel wise fashion with a 50% cross over between crops taken of observations, hence 125ms increments. This was inspired by Phinyomark et al. [11] who found that using a 250ms window size with 50% cross over yielded optimal results, although we also test for different window sizes in our results section.

Additionally, we also appeal to two simpler methods of just using raw values, rather than AR parameters, as feature vectors. As was done before, we augment our dataset before processing our data. We then make use of the window cropping method, also used above when aggregating AR parameters, first introduced by Liu et al.[10], where with a specified crop size, random crops of each signal are taken. However, since that augmentation method was made more generally for financial time series, it does not consider multiple channels. Hence, we made a slight adjustment and merged values across all channels into a single array, and random continuous crops of set length (windows) were taken from this array.

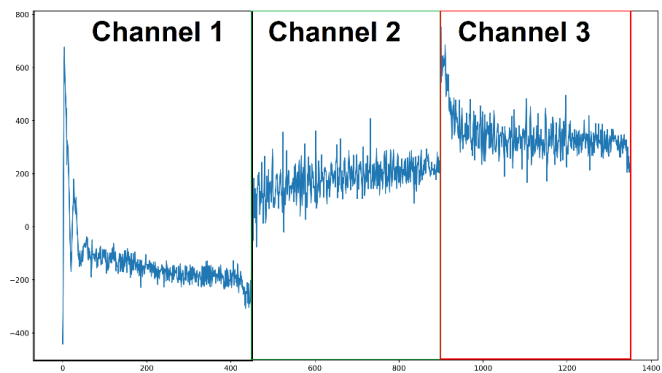

Fig 3. An example of an observation that was merged across channels.

The second type of dataset we use that consists of raw values, again involves augmenting the base dataset into 1500 observations, but now instead of merging the observation across channels into a singular array, windows of data are taken, with no cross over, across the 3 by 450 matrix representing each observation, particularly, if X is an observation:

$$X = \begin{bmatrix} X_1^{(1)} & \cdots & X_{450}^{(1)} \\ \vdots & \ddots & \vdots \\ X_1^{(3)} & \cdots & X_{450}^{(3)} \end{bmatrix}$$

An example of a window of this observation then taken in a channel wise fashion, if we were to use a window size of n $\leq$ 450, then this array would have the format:

$$[\ X_1^{(1)} \quad X_1^{(2)} \quad X_1^{(3)} \cdots X_n^{(1)} \ X_n^{(2)} \ X_n^{(3)}\ ]$$

The inspiration for taking crops across channels, rather than having each crop staying within its respective channel is to hopefully create dense representations of these windows, but with predictably sparse regions, so that when these feature vectors are passed into a model, they are as distinguishable from each other as possible if these sparse regions are to appear in a uniform fashion as we predict.

We then conclude this section with our model choice. For many devices used in HCI applications, there may be a limited memory, especially if one is trying to create a cost-effective system with high efficiency. For this reason, we move away from intense deep learning models, and instead move to tree-based algorithms, and attempt to choose features that can achieve a high accuracy model, but also while trying to keep our number of estimators low with the assumption that this model would need to run on any such device with limited memory. We already mentioned the popular choice of Random Forest in [6], but we also see the success of many XGBoost models generalizing quite well in more complex problem spaces, such as the identification of American Sign language hand gestures with 85% accuracy [12].

# III.    Results

*III.1 Results with Random Crops across array of Merged Channels*

We first present below a table using this feature extraction method using Random Forest and XGBoost models. The adjustable parameter for each data set is the size of the random crop, and we list next to each crop the maximum k cross-fold accuracy (k=5) achieved on a grid search with the associated data set, and the associated hyper parameters are listed as well. This convention is used for each method of feature extraction.

| Crop Size(ms) | Max. Depth | Num. Estimators | Max. k-CV Accuracy (%) |
|---|---|---|---|
| 1200 | 20 | 10 | 91.23 |
| 1100 | 20 | 10 | 89.5 |
| 1000 | 30 | 5 | 86.82 |
| 900 | 30 | 5 | 83.97 |
| 800 | 30 | 20 | 78.11 |
| 700 | 30 | 20 | 72.64 |
| 600 | 20 | 20 | 69.03 |
| 500 | 30 | 20 | 68.00 |
| 400 | 30 | 20 | 65.31 |
| 300 | 30 | 15 | 61.58 |

Table 1. Results using a Random Forest model and extracting features through random crops on merged channels of each signal.

| Crop Size(ms) | Max. Depth | Subsample | Col. Sample by Tree | Max. k-cv Accuracy (%) |
|---|---|---|---|---|
| 1200 | 10 | 0.8 | 0.6 | 94.11 |
| 1100 | 10 | 0.6 | 0.8 | 92.06 |
| 1000 | 20 | 0.8 | 0.6 | 89.47 |
| 900 | 20 | 0.8 | 0.6 | 88.42 |
| 800 | 20 | 0.8 | 0.6 | 85.71 |
| 700 | 20 | 0.6 | 0.8 | 84.52 |
| 600 | 10 | 0.8 | 0.6 | 80.66 |
| 500 | 20 | 0.6 | 0.6 | 76.98 |
| 400 | 20 | 0.8 | 0.8 | 73.70 |
| 300 | 20 | 0.6 | 0.8 | 69.49 |

Table 2. Results using an XGBoost model and extracting features through random crops on merged channels of each signal.

*III.2 Results using channel-wise crop of signal with no crossover*

| Crop Size(ms) | Max. Depth | Num. Estimators | Max. k-CV Accuracy (%) |
|---|---|---|---|
| 450 | 30 | 20 | 90.33 |
| 400 | 30 | 10 | 91.72 |
| 350 | 20 | 10 | 92.48 |
| 300 | 10 | 10 | 91.42 |
| 250 | 30 | 10 | 92.13 |
| 200 | 30 | 10 | 89.65 |
| 150 | 30 | 10 | 92.13 |

Table 3. Results using a Random Forest model and extracting features through channel wise crops with specified crop length and no cross over.

| Crop Size(ms) | Max. Depth | Subsample | Col. Sample by Tree | Max k-cv Accuracy (%) |
|---|---|---|---|---|
| 450 | 10 | 0.6 | 0.6 | 81.44 |
| 400 | 10 | 0.8 | 0.6 | 81.44 |
| 350 | 20 | 0.6 | 0.8 | 93.97 |
| 300 | 20 | 0.8 | 0.8 | 94.92 |
| 250 | 10 | 0.6 | 0.8 | 95.18 |
| 200 | 10 | 0.6 | 0.6 | 91.24 |
| 150 | 10 | 0.8 | 0.8 | 83.59 |

Table 4. Results using an XGBoost model and extracting features through channel wise crops with specified crop length and no cross over.

*III.3 Results using AR parameters of each respective channel merged into single feature vectors. Has a maximum crop size of 450ms since in one given channel for an observation, the number of elements in that channel is 450 elements*

| Crop Size(ms) | Max. Depth | Num. Estimators | Max. k-cv Accuracy (%) |
|---|---|---|---|
| 125 | 30 | 15 | 78.11 |
| 250 | 20 | 15 | 84.27 |
| 375 | 20 | 20 | 83.57 |
| 450 | 30 | 20 | 82.87 |

Table 5. Results using a Random Forest model and using AR parameters on channel-wise crops with specified crop length and 50% crossover

| Crop size(ms) | Max. Depth | Subsample | Col. Sample by Tree | Max. k-cv Accuracy(%) |
|---|---|---|---|---|
| 125 | 10 | 0.6 | 0.8 | 80.23 |
| 250 | 20 | 0.6 | 0.6 | 85.11 |
| 375 | 10 | 0.6 | 0.6 | 83.47 |
| 450 | 10 | 0.6 | 0.6 | 82.89 |

Table 6.  Results using an XGBoost model and using AR parameters on channel-wise crops  with specified crop length and 50% crossover

## IV.    Discussion and Conclusion

We have been able to create a high accuracy model with a  limited amount of training data using time series augmentation and cropping of our data, supporting the idea that high quality models with the aim of classifying bioelectrical signals can perform sufficiently well with only a couple of high-quality samples, only ten per class in our case. Additionally, towards the goal of being able to create computationally inexpensive models that can be ran on low memory electric devices, we shy away from neural network based approaches, and we can even see from our results section that in one instance in table 3, 91.42% k cross validation accuracy (k=5) was achieved with a relatively small model size of only 10 estimators, and a maximum depth of 10 per. In terms of how our models compare to others, we considered a baseline performance to be the 96.67% k cross validation accuracy (k=6), mentioned in Gokgoz et al.[6] that made use of AR parameters. We were able to record a maximum k cross-fold validation accuracy of 95.18% with the fifth model in table 4, supporting one of our main ideas that quality performance can also be reached with simpler methodologies.

With a small dataset and the same amount of augmentation, we see that using raw voltage values of an EMG signal unexpectedly performed better than using AR parameters. As was hinted in Phinyomark et. al when using both Random Forest and XGBoost models on our data, peak k cross-validation accuracy was achieved using a 250ms window with 50% crossover. No matter the method of feature extraction, XGBoost performed better than Random Forest, but again, we stress the importance that a model that is ready to be deployed on unseen data does not necessarily have the most impressive metrics on pre-built training and testing datasets.

There are others, such as Andrew Ng, that have taken on the belief that the "Big Data" methodology is not the solution going forward, and rather the use of small datasets can be used for complex problems in AI [17]. In the interest of producing artificial general intelligence in a manner that can mirror human cognition, it might be valuable for the field to work on model architectures that can make "abstractions" from fewer and fewer training instances in a human like manner, rather than the creation of models that can perform well on high amounts of data that only large corporations are able to acquire.

## Acknowledgements:

This research was partially supported by Triton NeuroTech X. We thank our colleague Alessandro D'Amico, a PhD student in Cognitive Science at U.C San Diego, who provided insight and expertise that greatly assisted this research this his role in collecting our EMG data, and performing filtering and smoothing of our data on a public code repository, although they may not agree with all of the interpretations/conclusions of this paper.